%% file: main.tex
\documentclass{article}

\usepackage{microtype}
\usepackage{graphicx}
\usepackage{subfigure}
\usepackage{booktabs} 

\usepackage{hyperref}

\usepackage[accepted]{icml2020}

\icmltitlerunning{Graph-based, Self-Supervised Program Repair from Diagnostic Feedback}

\usepackage{amsmath}
\usepackage{lipsum} 
\usepackage{algorithm}
\usepackage{setspace}

\makeatletter
\renewcommand\section{\@startsection{section}{1}{\z@}{-0.06in}{0.001in} {\large\bf\raggedright}}
\def\subsection{\@startsection{subsection}{2}{\z@}{-0.05in}{0.001in}{\normalsize\bf\raggedright}}
\renewcommand\subsubsection{\@startsection{subsubsection}{3}{\z@}{-0.04in}{0.0001in}  {\normalsize\bf\raggedright}}
\renewcommand\paragraph{\@startsection{paragraph}{4}{\z@}{0.001ex plus 0.001ex minus .001ex}{-1em}{\normalsize\bf}}
\makeatother

\usepackage{xcolor}
\usepackage{multirow}
\usepackage{makecell}
\usepackage{arydshln}
\renewcommand\ttdefault{cmvtt}

\newcommand{\erri}{k}

\begin{document}
\setlength{\abovedisplayskip}{4pt}
\setlength{\belowdisplayskip}{4pt}

\twocolumn[

\icmltitle{Graph-based, Self-Supervised Program Repair from Diagnostic Feedback}




\begin{icmlauthorlist}
\icmlauthor{Michihiro Yasunaga}{su}
\icmlauthor{~~Percy Liang}{su}
\icmlauthor{~~~~~~~}{}
\end{icmlauthorlist}

\icmlaffiliation{su}{Stanford University, Stanford, CA}

\icmlcorrespondingauthor{Michihiro Yasunaga}{myasu@cs.stanford.edu}

\icmlkeywords{Program repair, self-supervision, graph}

\vskip 0.3in
]



\printAffiliationsAndNotice{}  

\begin{abstract}\vspace{-0mm}
\begin{spacing}{1}
We consider the problem of learning to repair programs from diagnostic feedback (e.g., compiler error messages). Program repair is challenging for two reasons: First, it requires reasoning and tracking symbols across source code and diagnostic feedback. Second, labeled datasets available for program repair are relatively small.
In this work, we propose novel solutions to these two challenges. First, we introduce a program-feedback graph, which connects symbols relevant to program repair in source code and diagnostic feedback, and then apply a graph neural network on top to model the reasoning process. Second, we present a self-supervised learning paradigm for program repair that leverages unlabeled programs available online to create a large amount of extra program repair examples, which we use to pre-train our models.
We evaluate our proposed approach on two applications: correcting introductory programming assignments (DeepFix dataset) and correcting the outputs of program synthesis (SPoC dataset). Our final system, DrRepair, significantly outperforms prior work, achieving 68.2\% full repair rate on DeepFix (+22.9\% over the prior best), and 48.4\% synthesis success rate on SPoC (+3.7\% over the prior best).
\end{spacing}
\vspace{-4mm}
\end{abstract}

\section{Introduction}

\begin{figure}[!t]
    \vspace{-2mm}
    \hspace{-3mm}
    \includegraphics[width=0.504\textwidth]{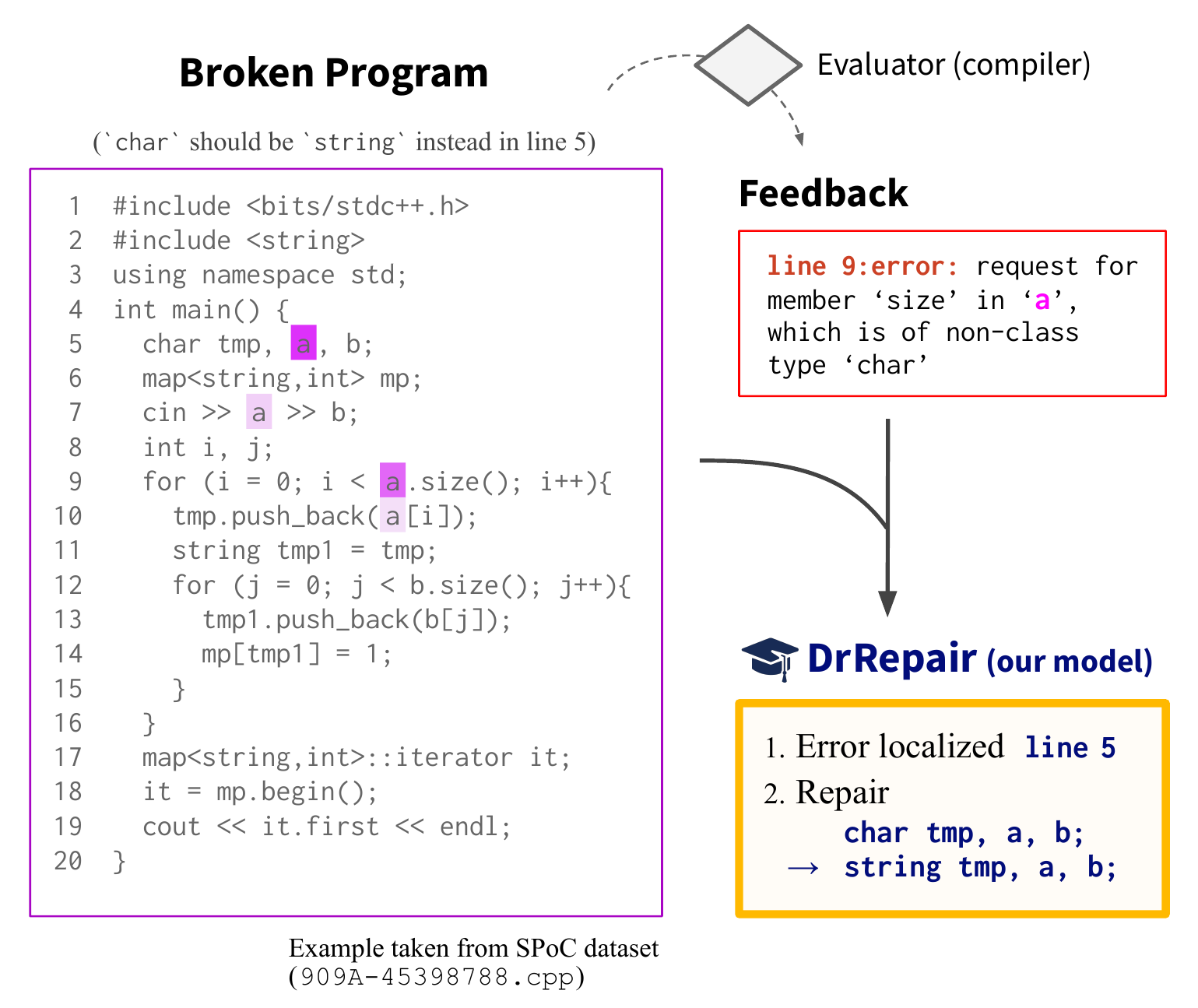}\vspace{-8mm}
    \caption{
    Given a broken program and diagnostic feedback (compiler error message), our goal is to localize an erroneous line and generate a repaired line.
    }\vspace{-4mm}
  \label{fig:task}
\end{figure}

Automatic program repair has the potential to dramatically improve the productivity of programming.
In particular, a common source of program errors are compiler errors, which include
use of unresolved symbols, missing delimiters (e.g. braces), and type errors. 
These errors are commonly observed in both beginner programmers \cite{Automatic_grading} and professional developers
\citep{google_case},
as well as in the predicted code of program synthesis \citep{spoc2019}.
Accordingly, the use of machine learning in fixing compiler errors has garnered significant interest recently \citep{DeepFix,SampleFix,DeepDelta}.

In this work, we consider the problem of learning to repair programs based on diagnostic feedback (compiler error messages).
Figure \ref{fig:task} illustrates the setup. Given a broken program and diagnostic feedback, we aim to localize an erroneous line in the program and generate a repaired line. 
Learning program repair has two major challenges: 
First, the system needs to connect and {\it jointly} reason over the broken source code and the diagnostic feedback \citep{fitzgerald2008debugging}. 
Second, existing works rely on manual effort to curate labeled datasets for program repair (e.g. $\langle$broken program, fixed program$\rangle$ pairs), which does not scale up \citep{DeepDelta}.
Here we present \textit{DrRepair}, a novel approach to program repair that addresses these two challenges. 
Our key innovations are two-fold: 
1) modeling of program repair with \textit{program-feedback graphs} and 2) self-supervised learning with unlabeled programs.

\begin{figure}[!t]
    \vspace{-2mm}
    \includegraphics[width=0.46\textwidth]{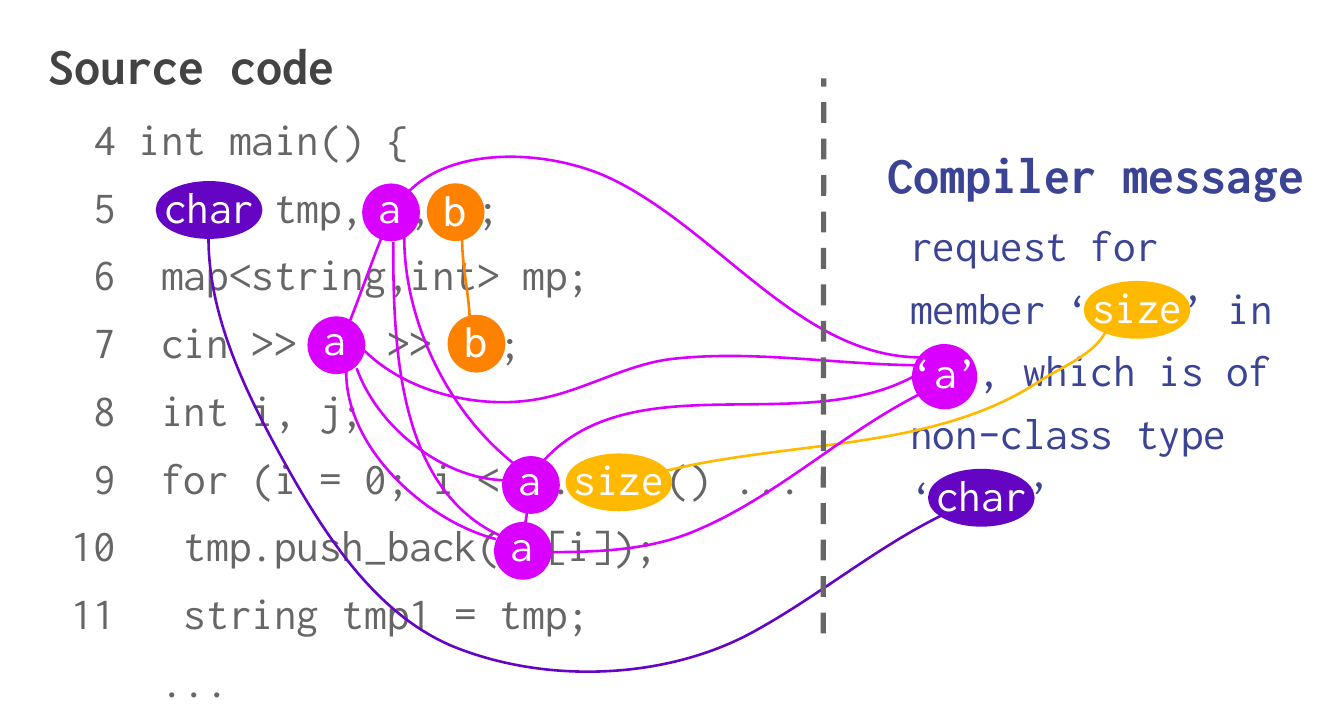}\vspace{-3mm}
    \caption{Illustration of \scalebox{0.93}[1]{\textbf{program-feedback graph}, corresponding} to the 
    example in Fig. \!\ref{fig:task}. The graph captures long-range dependencies of symbols to help model the reasoning process of program repair.}
    \label{fig:graph}\vspace{-4mm}
\end{figure}

\textit{Program-feedback graph}.
Program repair requires 
reasoning jointly over the symbols (e.g. identifiers, types, operators) across source code and diagnostic feedback.
For instance, in the example given in Fig.\,\ref{fig:task}, while the compiler message points to line 9, the error is related to the type of identifier `\texttt{a}', and one needs to track how `\texttt{a}' has been used or declared earlier to resolve this error. 
To formalize this reasoning process, we
propose a joint graph representation of a program and diagnostic feedback that captures the underlying semantic structure of symbols in the context of program repair ({program-feedback graph}). 
Specifically, it takes all identifiers (e.g. \texttt{a}, \texttt{b}) in the source code and any symbols in the diagnostic arguments (e.g. ‘\texttt{a}’, ‘\texttt{char}’) as nodes, and connects instances of the same symbols with edges to encode the semantic correspondence (Fig.\,\ref{fig:graph}). 
We then design a neural net model with a graph-attention mechanism \citep{velikovi2017graph} on the program-feedback graph to model the symbol tracking process described above.
While prior works in program repair purely apply sequence-to-sequence (seq2seq) models to programs \citep{DeepFix,SampleFix} or rely on the program's Abstract Syntax Tree (AST) representations \citep{DeepDelta,Graph2Diff}, our program-feedback graph directly connects symbols involved in the reasoning process of program repair, and allows efficient information flow across them.

\textit{Self-supervised learning.}
Motivated by the vast amount of program data  available online (e.g. GitHub  has 28 million public repositories), we propose a self-supervised learning paradigm for program repair that leverages unlabeled programs to create a large amount of {extra} training data. 
Specifically, we collect working programs from online resources related to our problem domain (programming contests in our case), and design a procedure that corrupts a working program into a broken one, thereby generating new examples of $\langle$broken program, fixed program$\rangle$. 
In our experiments, we prepare extra data \scalebox{0.7}[0.9]{$\sim$}10 times the size of original datasets in this way, use it to pre-train our models, and fine-tune on the target task. 
We also describe an effective corruption procedure that covers a diverse set of errors.
While prior works in program repair rely on labeled datasets \cite{DeepDelta,Graph2Diff,spoc2019}, we are the first to present a self-supervised learning method for program repair that leverages unlabeled programs online.

We evaluate the efficacy of our proposed approach on two applications, using publicly available datasets:\vspace{-4mm}
\begin{itemize}
    \setlength{\leftskip}{-2mm}
    \setlength{\itemsep}{-1mm}
    \item[{a)}] Correcting introductory programming assignments. We use DeepFix dataset \citep{DeepFix}, where the task is to repair 
    broken 
    C programs submitted by 
    students.
    
    \item[{b)}] Correcting the output code in program synthesis. We use the
    SPoC dataset \citep{spoc2019}, where the task is to translate pseudocode into C++ implementation, and programs synthesized by prior models (seq2seq) often fail to compile. 
    We apply our repair model to correct the candidate programs generated in this task. \vspace{-4mm}
\end{itemize}
Experimental results show that our approach (DrRepair) outperforms prior work significantly, achieving 68.2\% full repair on the DeepFix test set ({+22.9\%} absolute over the prior best), and 48.4\% synthesis success rate on the SPoC test set
(+3.7\% absolute over the prior best at the time of this work).
Additionally, our analysis shows that the use of a program-feedback graph is particularly helpful for fixing errors that require reasoning over multiple lines of code, and that self-supervised pre-training facilitates the learning of program repair for the types of errors with fewer training examples in the original dataset.

\begin{figure*}[!t]
    \vspace{-2mm}
    \hspace{-1mm}
    \centering 
    \includegraphics[width=0.99\textwidth]{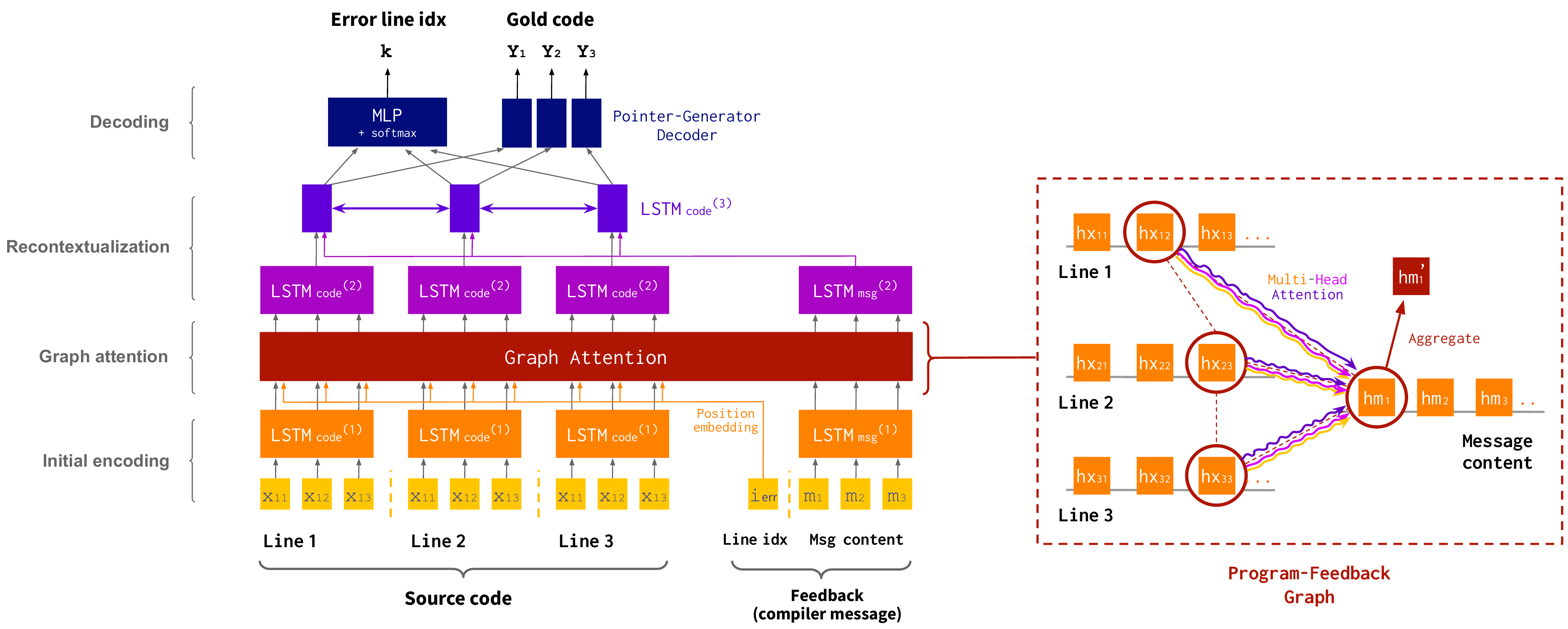}
    \vspace{-4mm}
    \caption{DrRepair model. It takes in a program $x\!=\!(x_1, ... , x_L)$ and diagnostic feedback from a compiler $f \!=\! (i_\text{err}, m_\text{err})$ as inputs (\textbf{bottom}), encodes them via LSTM and graph attention layers, and decodes the error line index $\erri$ and repaired code $y_{\erri}$ (\textbf{top}). The right-hand side illustrates the graph attention mechanism.
    Best viewed in color.
    }
    \vspace{-3mm}
\label{fig:model}
\end{figure*}

\section{Problem statement}

Figure \ref{fig:task} illustrates the program repair task. 
The system is given (a) a broken program with $L$ lines, $x \!=\! (x_1, ..., x_L)$, 
and (b) diagnostic feedback provided by a compiler, $f \!=\! (i_{\rm{err}}, m_{\rm{err}})$, 
where $i_{\rm{err}}$ denotes the reported line number, and $m_{\rm{err}}$ the error message (a sequence of tokens).
If the compiler returns multiple error messages, we use only the first one.\footnote{Note that here we are defining a module that repairs a single line of code in a program.
We describe how we apply this repair module to programs with multiple errors in \S \ref{sec:experiments}. We also explain the application-dependent evaluation metrics in \S \ref{sec:experiments}.}  
Our task is to identify the index of an erroneous line $\erri \in \{1, \dots, L\}$ (\textit{error localization}), and generate a repaired version of the line $y_\erri$ (\textit{repair}).
Let $y\!=\!y_{1:L}$ denote the fixed version of the full program ($y_i = x_i$ for $i \neq \erri$).
In the example given in Figure \ref{fig:task}, 
$x_5 \!=\!$ ``\,\texttt{char tmp, a, b;}\,'',~
$i_{\rm{err}} \!=\! 9$, ~$m_{\rm{err}} \!=\!$ ``\,\texttt{request for \!...\! type ‘char’}\,'', and
$\erri \!=\! 5$, ~$y_\erri \!=\!$ ``\,\texttt{string tmp, a, b;}\,''.
Note that the line number reported by a compiler ($i_{\rm{err}}$) does not necessarily match the line we need to repair ($\erri$).

\section{Approach}

We approach program repair from two angles.
First, we propose a \textit{program-feedback graph} to model the reasoning process involved in program repair.
Second, 
we introduce a self-supervised learning paradigm 
that leverages unlabeled programs to create a large amount of extra training data.

\subsection{Modeling}
To model program repair,
we start off with a sequence-to-sequence learning setup, and
incorporate the information of a program-feedback graph
through a graph attention model, 
which we describe below.
Given an input program $x_{1:L}$ and its feedback $f \!=\! (i_\text{err}, m_\text{err})$, we first tokenize each line $x_i$ and the compiler message $m_\text{err}$ into a sequence of symbols: $x_i \!=\! (x_{i1}, x_{i2}, ... )$ and $m_\text{err} \!=\! (m_{1}, m_{2}, ... )$.
As seen in our motivating example in Fig.\,\ref{fig:task}, program repair requires reasoning and tracking symbols 
across different lines of code and compiler messages
(e.g., given the compiler error about `\texttt{a}', a programmer will jump to the source code line reported by the message, and then track how `\texttt{a}' has been used \!/\! declared in earlier lines).
These long-range dependencies of tokens are difficult to capture using previous seq2seq or AST-based models, which only propagate information locally at the line or syntax level \cite{DeepFix,DeepDelta}.
To enable more efficient information flow,
we introduce a program-feedback graph $G$ that \textit{directly} connects tokens relevant to the reasoning of program repair.

\subsubsection{\rm{\textbf{Program-feedback graph}}}

A program-feedback graph
$G\!=\!(V, E)$ has nodes 
$V$ that consist of tokens in the diagnostic arguments (those within `\,' in the message, i.e., \texttt{size}, \texttt{a}, \texttt{char} in Fig.\,\ref{fig:graph}), their occurrences in the source code, and all remaining identifiers in the code (e.g. \texttt{a}, \texttt{b}, \texttt{i}, \texttt{j}). 
The type of each token,
such as identifier (for \texttt{a}), operator (for \texttt{=}) and data type (for \texttt{char}), is recognized by the C/C++ tokenizer in \citet{DeepFix}. 

We then form the graph by connecting identical tokens in $V$ with undirected edges ($E$) to capture the semantic correspondence.
The resulting graph is as a set of cliques, one for each symbol (e.g. `\texttt{a}').
We keep the program-feedback graph simple
for two reasons:
1) we use the graph and graph-attention to specifically capture the (long-range) connections of tokens crucial to program repair reasoning, and perform other local information propagation via LSTMs (we elaborate in \S \ref{sec:approach-model-architecture}), and
2) it is nontrivial to analyze the code further (e.g. parsing) to add information to the graph, as the program can be syntactically ill-formed. Compared to AST-based graph representations \citep{learning_to_Allamanis18,Graph2Diff},
our program-feedback graph is more relaxed and robust to errors in source code.

\renewcommand\ttdefault{cmvtt}

\begin{table*}[!t]
    \vspace{-3mm}
    \hspace{1mm}
    \centering 
    \includegraphics[width=0.975\textwidth]{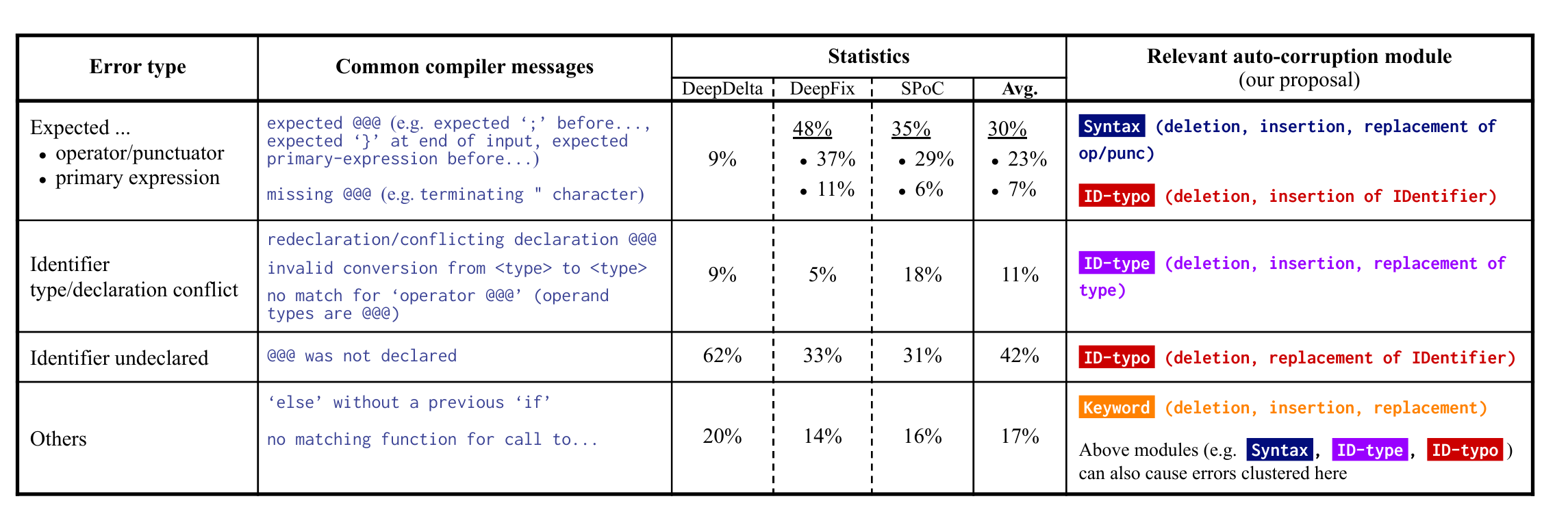}
    \vspace{-6.5mm}
    \caption{Analysis of common compiler errors in three settings: experienced developers \scalebox{0.95}{(DeepDelta)}, beginner programmers  \scalebox{0.95}{(DeepFix)}, and predicted code of program synthesis \scalebox{0.95}{(SPoC)}. For \scalebox{0.95}{DeepDelta}, the statistics is taken from \citet{DeepDelta}.
    The rightmost column shows the program perturbation modules that we design to generate corresponding types of errors.
    }
    \vspace{-3mm}
\label{tbl:err_stats}
\end{table*}

\subsubsection{\rm{\textbf{Model architecture}}}
\label{sec:approach-model-architecture}
Fig.\,\ref{fig:model} illustrates 
our program repair model. It has an encoder that takes in a program $x$ and feedback $f$, and a decoder that predicts a distribution over which line is erroneous $\erri$ and a repaired line $y_\erri$.
The encoder has three stages: 1) initial encoding $\mathbf{h} \!=\! \text{InitEnc}(x, f)$ which
encodes each input token at the line level,
2) graph attention $\mathbf{g} \!=\! \text{GraphAttn}(\mathbf{h})$ which propagates information across tokens on a program-feedback graph, and 
3) recontextualization $\mathbf{s} \!=\! \text{ReContext}(\mathbf{g})$ which contextualizes token representations at the line level again to produce an embedding $\mathbf{s}_i$ for each line $i$.
Finally, $\text{Decode}(\mathbf{s})$
outputs a distribution over the erroneous line index and a repaired line $(\erri, y_\erri)$.
We describe each of the model stages in detail below.

\newcommand\lstmname[2]{\scalebox{0.95}{$\text{LSTM}_{#1}^{\scalebox{0.7}{\,#2}}$}}

\textbf{Initial encoding.~~}
Given source code $x_{1:L}$ and feedback $f\!=\!(i_\text{err}, m_\text{err})$ (Fig.\,\ref{fig:model} {bottom}), we 
encode each line $x_i$ and compiler message $m_\text{err}$ with two bidirectional LSTM networks \citep{Hochreiter:1997:LSM:1246443.1246450}, 
\lstmname{\text{code}}{(1)}
and \lstmname{\text{msg}}{(1)}.
For the tokens in the source code, 
we also inject the information of the reported line index ($i_\text{err}$) by concatenating the outputs of \lstmname{\text{code}}{(1)} with the positional encoding \citep{vaswani2017attention} of the line offset $\Delta i \!=\! i_\text{err} \!-\!i$, and applying a feedforward network.
We denote the representation of each token in the code and message at this point as $\mathbf{h}_{x_{ij}}$ and $\mathbf{h}_{m_\ell}$, respectively.
This stage is analogous to the input encoding in \citet{spoc2019}.

\textbf{Graph attention.~~}
Next, to model the reasoning (symbol tracking) process in program repair, we use a graph attention network \citep{velikovi2017graph} to allow information to flow across symbols in the program-feedback graph $G$ (Fig. \!\ref{fig:model} {right}). 
In a $N$-layer graph attention network, each layer computes contextualized representations of tokens via
\begin{align}
     {\mathbf{c}^{n}} &= 
     \text{{Attention}}_{{\,G}}
     (\mathbf{h}^{n-1}) \\
    \mathbf{h}^{n} &= \text{MLP}([\mathbf{h}^{n-1}; \mathbf{c}^{n}])
\end{align}
where $\mathbf{h}^{n-1}$, $\mathbf{h}^{n}$ denote the input \!/\! output representation of each token at the $n$-th layer. Initially, $\mathbf{h}^{0}$ is  $\mathbf{h}_{x_{ij}}$ or $\mathbf{h}_{m_\ell}$, and the final output $\mathbf{g} \!=\! \mathbf{h}^{N}$.
\text{Attention}$_{{\,G}}(\mathbf{h}_t)$ 
computes attention weights over the neighboring nodes of a token $t$ on the graph $G$, $\mathcal{N}_G(t)$, and takes the weighted average of the token representations among $\mathcal{N}_G(t)$. 
\text{MLP} is a feedforward network.
For a more detailed description about graph attention, we refer readers to \citet{velikovi2017graph}.

\textbf{Recontextualization.~~}
We allow the information updated via the graph to propagate on the local context again, by 
passing the token representations $\mathbf{g}$ to additional sequence networks, \lstmname{\text{code}}{(2)} and \lstmname{\text{msg}}{(2)}.
We obtain an {embedding} of each line $i$ by concatenating their final hidden states,
\begin{align}
    \mathbf{r}_i = \bigl[ \lstmname{\text{code}}{(2)}(\mathbf{g}_{x_{i\cdot}})^{\text{final}};~ \lstmname{\text{msg}}{(2)}(\mathbf{g}_{m_{\cdot}})^{\text{final}} \bigr]
\end{align}
which is further contextualized
to be the final line embedding $\mathbf{s}_i$, via $\mathbf{s}_{1:L} = \lstmname{\text{code}}{(3)} (\mathbf{r}_{1:L})$ (Fig.\,\ref{fig:model} {top}).

\textbf{Decoding.~~}
Given the line embeddings $\mathbf{s}_{1:L}$,
we model the probability of a line $\erri \in \{ 1, \dots, L \}$ being erroneous via a feedforward network,
and model its repair $y_\erri$, via a pointer-generator decoder \citep{See_2017}:
\begin{align}
    \label{eq:err-idx}
    p(\erri \,|\, \mathbf{s}_{1:L}) & = \text{softmax}(\text{MLP}(\mathbf{s}_{1:L}))\\
    \label{eq:ptr-gen}
    p(y_\erri \,|\, \mathbf{s}_{1:L}) & = \text{PtrGen}(\mathbf{s}_\erri).
\end{align}

\textbf{Training.~~}
A training example consists of a broken program $x$, feedback $f$, an erroneous line index $\erri$,
and the repaired line $y_\erri$.
The loss on a given example is the standard
negative log-likelihood, $-\log p(\erri, y_\erri \mid x, f)$.
The error localization and repair components are learned jointly.
In \S \ref{sec:approach-selfsuper} and \S \ref{sec:exp-setup}, we will discuss how we generate training examples of this form for pre-training and target applications.

\begin{table}[!t]
    \vspace{-1mm}
    \includegraphics[width=0.49\textwidth]{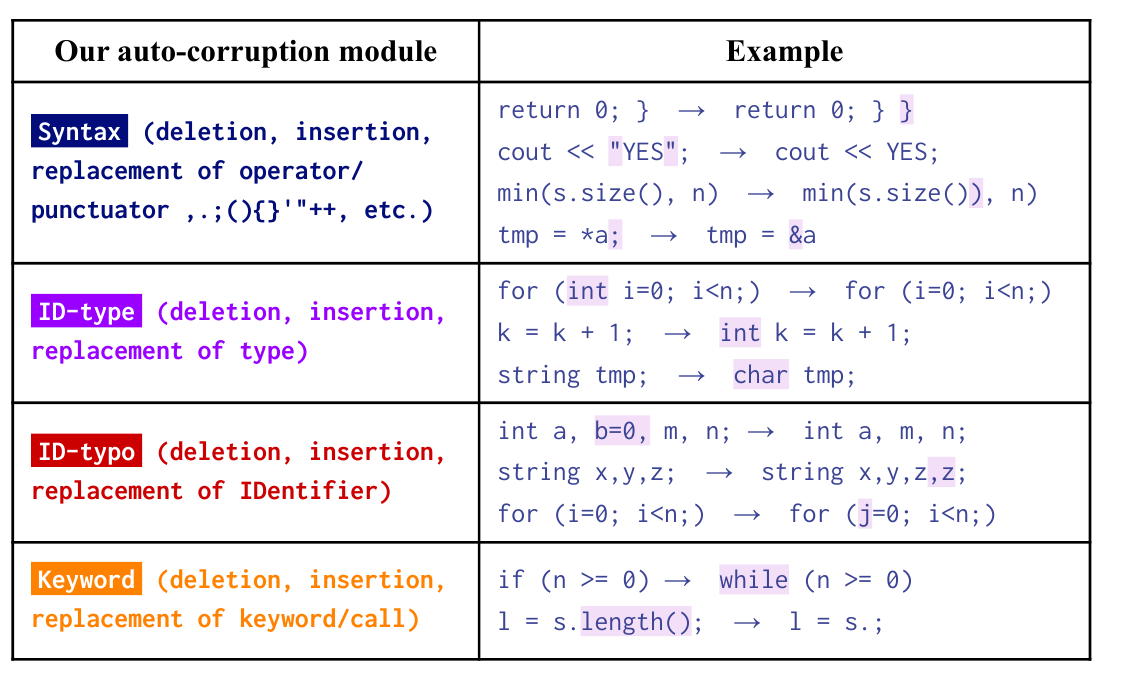}
    \vspace{-10mm}
    \caption{Proposed program perturbation modules for generating self-supervised data.
    }
    \vspace{-5mm}
\label{tbl:corrupt_ex}
\end{table}

\subsection{Self-supervised learning}
\label{sec:approach-selfsuper}

\renewcommand\ttdefault{cmtt}
Labeled datasets for program repair ($\langle x,y \rangle$ pairs) are limited in size (10--100K data points) \citep{DeepDelta}, 
but there is a vast amount of unlabeled programs available online: for instance,
GitHub\footnote{\scalebox{0.9}{\url{https://github.com/}}} 
alone has 28 million public repositories as of 2019.
Can we leverage this freely-available code to improve the learning of program repair? 
\renewcommand\ttdefault{cmvtt}

With this motivation, 
we propose a new self-supervised learning paradigm that utilizes unlabeled, working programs to create a large amount of training data for program repair.
Specifically, we first collect a large set of working programs $y$'s (ones that compile, in our setting), related to the domain of interest.
We design a randomized procedure $\mathcal{P}$ that automatically corrupts $y$ into a broken program $x$ to generate a new training example $\langle$broken code $x$, ground-truth $y\rangle$.
We repeatedly apply this procedure to the collected programs,
and use the generated training data to perform pre-training \citep{Erhan_pretrain} of our model, facilitating it to learn useful representations for program repair (\textit{self-supervised pre-training}).
Later, we fine-tune the model on a labeled, original (in-domain) dataset.

Below, we describe the details of our program corruption and data generation process.

\paragraph{Program corruption procedure.}

To design an effective corruption procedure that covers a diverse set of program errors, we first analyzed common compiler errors in three settings: experienced developers, beginner programmers, and predicted code of program synthesis.
\renewcommand\ttdefault{cmtt}
For each case, we collected statistics from \citet{DeepDelta}, DeepFix dataset \citep{DeepFix} and SPoC dataset \citep{spoc2019}, and grouped the errors into four major categories:
``{{Expected \!...}}'', ``{{Type/declaration conflict}}'', ``{{Identifier undeclared}}'', and ``{{Others}}'' (details in Table \ref{tbl:err_stats}). 

Motivated by this analysis, we design a set of perturbation modules (heuristics), $\mathcal{M}$, that aim to modify source code to cause the above types of errors.
Specifically, $\mathcal{M}$ consists of
\vspace{-3.5mm}
\begin{itemize}
\setlength{\leftskip}{-3mm}
\setlength{\itemsep}{-0.4mm}
    \renewcommand\ttdefault{cmtt}
    \item {\textbf{Syntax}}, which randomly deletes, inserts or replaces an operator \!/\! punctuation, such as \texttt{,.;()\string{\string}[]"++<<}. This module causes various errors such as ``\texttt{expected @@@}''.
    
    \item {\textbf{ID-type}}, which randomly deletes, inserts or replaces an identifier (ID) type such as \texttt{int}, \texttt{float}, \texttt{char}.
    This causes errors such as conflicting types and redeclaration.
    
    \item {\textbf{ID-typo}}, which randomly deletes, inserts or replaces an identifier. 
    This module causes errors such as missing primary expressions and 
    undeclared identifiers.
    
    \item {\textbf{Keyword}}, which randomly deletes, inserts or replaces a use of program language keyword or library function, such as \texttt{if} and \texttt{size()}.
    This module can cause other miscellaneous errors.
    \vspace{-3.5mm}
\end{itemize}
Table \ref{tbl:corrupt_ex} provides concrete examples of each
module.
Note that each module makes a single change to source code at a time.
Given the perturbation modules $\mathcal{M}$,
our program corruption procedure (named \textit{DrPerturb}) samples 1--5 modules from $\mathcal{M}$ (with replacement) and applies them to an input program sequentially.
We sample each module with probability 0.3, 0.5, 0.15, 0.05, respectively,
motivated by the distribution of errors found in our analysis (Table \ref{tbl:err_stats}).

We will show in our experiments that DrPerturb is significantly more effective than baseline corruption procedures such as randomly deleting tokens.

\paragraph{Data preparation details.}

As the program domain in our applications (DeepFix, SPoC) is C/C++ implementation of introductory algorithms, 
we turn to programs available on \url{codeforces.com}, which contains C++ code submitted by programming contest participants.
We collect accepted programs and filter out outliers (e.g. those longer than 100 lines),
following the procedure in \citet{spoc2019}. 
This yields 310K C++ programs that compile successfully, which is roughly 10 times the size of the original training data available in our applications (37,415 programs in DeepFix, 14,784 in SPoC).
For each program, we then create roughly 50 corrupted versions by applying DrPerturb and keeping ones that fail to compile. This yields $\sim$1.5M extra training examples of $\langle$broken code $x$, feedback $f$, correct code $y\rangle$, which we use to pre-train our program repair model.

Note that the collected program data share the same source with SPoC (\url{codeforces.com}),\footnote{We made sure that the programs collected for pre-training do not overlap with the exact programs in SPoC test sets.} but not exactly with DeepFix, which is C programming assignments. Nevertheless, we find
that the collected data is highly effective in both tasks, which we elaborate on in \S \ref{sec:experiments}.

\section{Experiments}
\label{sec:experiments}
We conduct an extensive evaluation of our approach via two applications: DeepFix\footnote{\scalebox{0.9}{\url{https://bitbucket.org/iiscseal/deepfix}}\vspace{-1mm}} \citep{DeepFix} and SPoC\footnote{\scalebox{0.9}{\url{https://sumith1896.github.io/spoc}}} \citep{spoc2019}, which are recent benchmarks for program repair and program synthesis, respectively.

\input{tbl_eval_deepfix_v2.tex}
\input{tbl_eval_spoc_dev.tex}

\subsection{Experimental setup}
\label{sec:exp-setup}
We summarize the setup of DeepFix and SPoC, and describe how we apply our program repair model to those tasks.

\subsubsection{DeepFix}
\textbf{Task.~~}
\label{sec:exp-setup-deepfix}
The DeepFix dataset contains C programs submitted by students in an introductory programming course, of which 37,415 are correct (compile) and 6,971 are broken (do not compile).
The average program length is 25 lines.
The broken programs are called \textit{raw test set} and may contain multiple errors.
The task is to repair them into ones that compile (\textit{full repair}; evaluation metric is full repair rate).

\textbf{Data processing.~~}
To generate training \!/\! dev data for repair models, we corrupt the correct programs in DeepFix using DrPerturb.
We call this the \textit{synthetic} data, as apposed to the raw test set.
We also call this the \textit{original} train \!/\! dev data to distinguish it from the \textit{extra} data prepared for pre-training, which is not exactly in the same domain as DeepFix.

\textbf{How to apply the repair model.~}
At test time, 
as the broken programs may contain errors in multiple lines, we apply the repair model iteratively until the program compiles or we reach the attempt limit of 5, as in \citet{DeepFix}.

\subsubsection{SPoC}
\label{sec:exp-setup-spoc}

\textbf{Task.~~}
The SPoC dataset consists of 18,356 C++ programs (avg. \!length 15 lines) 
collected from \url{codeforces.com}.
For each program $t \!=\! t_{1:L}$ (with $L$ lines), 
every line of code is annotated with natural language pseudocode, $s_{1:L}$. 
The task is to synthesize the target program $t$ from pseudocode $s$ within a budget of $B$ attempts (search iterations).
Prior work \citep{spoc2019} uses a seq2seq translation system to map each pseudocode line $s_i$ into a set of 100 candidate code pieces $\mathcal{C}_i \!=\! \{t_{ic_i} \,|\, c_i \!\in \![100] \}$, where candidate piece $t_{ic_i}$ has probability $p_{ic_i}$.
A full candidate program $t$ is a concatenation of candidate code pieces, and has score $p(t)$:\vspace{-0.5mm}
\begin{align}
    t = \texttt{concat}_{i=1}^L t_{ic_i}, ~~~~ p(t) = \scalebox{1}{\text{$\prod_{i=1}^L$}}~ p_{ic_i}, \vspace{-0.5mm}
\end{align}
where $c_i$ is to be searched for each line $i$.
\citet{spoc2019} then considers various search algorithms (e.g. best first search using the score) to efficiently find the correct program $t$ from this space of candidates.

\textbf{Why \& how to apply the repair model.~~}
As \citet{spoc2019} observed, the top candidates produced by this scoring metric exhibited syntactic or semantic incoherence (e.g. conflicting types) and fail to compile, because each candidate score $p_{ic_i}$ is calculated by line-level translation of pseudocode, ignoring the global context.
To address this issue, \citet{spoc2019} combined best first search with error localization; here we propose a search algorithm that also follows best first search, but attempts to \textit{repair} the current candidate program with our repair model {if} it does not compile, and adds the repaired code piece $t_{ic_i'}$ into the pool of candidate code pieces $\mathcal{C}_i$, with an updated score $p_{ic_i'}$.

\textbf{Data processing.~~}
We follow the data splits in \citet{spoc2019}, which consists of Train, Dev, TestP, and TestW.
We use TestP \!/\! TestW for the final evaluation of program synthesis, and use Train \!/\! Dev to train or validate our repair model.
To prepare train \!/\! dev data for the repair model, for each program $y \!=\! y_{1:L}$ in SPoC, we sample an error line index $k$ and substitute line 
$y_{\erri}$ with a candidate 
$c_{\erri j} \!\in\! \mathcal{C}_{\erri}$ generated from pseudocode line 
$s_{\erri}$. We then collect any modified program 
$y'$ that produces a compiler error
$f$. 
We call this \textit{original} train \!/\! dev data, to distinguish with the \textit{extra} data prepared for pre-training.

\input{tbl_eval_spoc_test.tex}

\subsection{Hyperparameters \& training details}
We set the dimension of input token embeddings and position embeddings to be 200 and 100. 
The LSTMs and graph attention networks have a state size of 200.
We use 3, 2, 1 and 2 layers for LSTM$^{\scalebox{0.7}{\,(1)}}$, graph attention net, LSTM$^{\scalebox{0.7}{\,(2)}}$ and LSTM$^{\scalebox{0.7}{\,(3)}}$, respectively, with dropout rate
0.3 applied to each layer \cite{JMLR:v15:srivastava14a}. 
The parameters of the models are optimized by Adam \cite{kingma2015adam}, with batch size 25, learning rate 0.0001, and gradient clipping 1.0 \cite{Pascanu2012}, on a GPU (GTX Titan X).

\subsection{Results}
\label{sec:results}

We describe our main results on DeepFix and SPoC here.
We use ``base'' to denote the version of our model that replaces graph attention with line-level LSTM layers (a pure sequence model), and 
``base \!+\! graph'' the one with graph attention. We train these models on the \textit{original} data (from DeepFix \!/\! SPoC) only.
``base \!+\! graph \!+\! pre-train'' denotes a ``base \!+\! graph'' model that is pre-trained with self-supervision on the \textit{extra} data and fine-tuned on the \textit{original} data.

\textbf{DeepFix.~~}
Table \ref{tbl:deepfix-eval} describes the performance of our repair model along with prior work. 
``Single Repair'' column shows the accuracy of repairing a single line (single step) on the \textit{synthetic} dev set, and
``Full Repair'' column shows the full repair acc. on the \textit{raw} test set, where our repair model is run iteratively (\S \ref{sec:exp-setup-deepfix}).
First, we observe that our base model (``our base'' row) achieves 62.5\% full repair acc., which outperforms prior works (those above the dashed line) by 15\% absolute.
We hypothesize that this is because our model uses compiler messages as input, but prior works do not (they consider direct mapping of broken code into its fix). 
To understand the importance of using compiler messages for program repair, we experimented with a version of our model that does not use compiler messages (``our base, no compiler message'').
We find that while it attains comparable scores to ``our base'' on the \textit{synthetic} dev, 
the performance drops a lot 
on the raw test set: 34.0\% acc., similar to the prior work (30--40\% acc.). This suggests that diagnostic feedback offered by compiler messages plays a crucial role in learning program repair, and without it, the model tends to learn superficial patterns present in the \textit{synthetic} train \!/\! dev data (hence the high scores on dev).

Next, we find that
our program-feedback graph (``base + graph'') provides a 3\% boost over ``base'' in full repair rate, and
self-supervised pre-training (``base + graph + pre-train'') provides a further improvement of 2\%, suggesting that
both the program-feedback graph and self-supervision provide complementary improvements.
Consequently, 
with the use of compiler messages, graph and pre-training,
our full system DrRepair (``base + graph + pre-train'') improves on the prior best (SampleFix) by 22.9\% in total, achieving a state-of-the-art result of 68.2\% full repair rate.

\input{tbl_perturb_analysis.tex}

\textbf{SPoC.~~}
Similar to DeepFix, we measure the single step repair accuracy on the SPoC dev set (Table \ref{tbl:spoc-dev}). 
The use of graph and pre-training both improve the repair performance (4.4\% and 3.2\% respectively; first three rows).
As the SPoC task contains pseudocode, 
we also experimented with a version of our repair model that takes in pseudocode as input in addition to the broken code and compiler message.
This provides a further boost in performance, achieving 68\% single repair acc. on SPoC dev set (bottom row).

We then apply our repair model to the program synthesis setting (TestP, TestW), as described in
\S \ref{sec:exp-setup-spoc}.
As seen in 
Table \ref{tbl:spoc-test},
our synthesis method
equipped with DrRepair (bottom row) improves on the best first search significantly (e.g. +6\% on TestP \!/\! TestW budget 100), 
suggesting that our repair model is useful for  program synthesis as well.

We note that 
a concurrent work \citep{semantic_scaffold} uses semantic constraints of programs to improve the search and achieves state-of-the-art results (46.1\% \!/\! 62.8\% on TestP \!/\! TestW). In contrast,
our method only requires blackbox access to a compiler or executor. We believe the two approaches are complementary and it would be interesting to combine the approaches.

\textbf{Example \& Visualization.~~}
Figure \ref{fig:task} gives a real example of the output of ``base+graph'', as well as 
visualization of graph attention, where the pink highlighting in the source code shows the computed attention weights w.r.t. the `\textcolor{red}{\texttt{a}}' in the compiler message (the darker, the higher).
It indicates that the model attends not only to the line reported by compiler (line 9), but also to the line that declared `{\texttt{a}}', which is the source of the error.
This way, we can interpret the reasoning performed by our repair model.

\input{tbl_graph_analysis.tex}

\begin{table*}[!t]
    \vspace{-1mm}
    \centering 
    \includegraphics[width=0.84\textwidth]{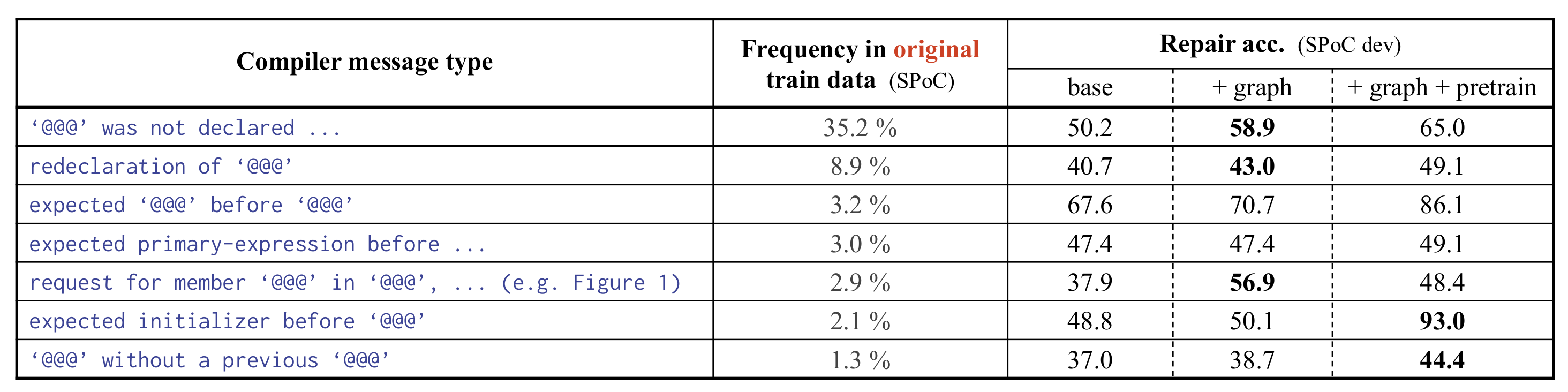}
    \vspace{-5mm}
    \caption{Breakdown of major compiler errors seen in SPoC dev (left), and the corresponding repair accuracy by our model variants (right).
    \textbf{Bold score} indicates a particularly big improvement from ``base'' to ``+ graph'' or from ``+ graph'' to ``+ graph + pretrain''. 
    }
    \vspace{-3mm}
\label{tbl:acc_vs_err}
\end{table*}

\subsection{Analysis}
We aim to understand 1) the effect of different program corruption procedures and 2) graph representation methods, and 3) when self-supervision or graph is useful.

\textbf{Different program corruption procedures.~~}
We compare DrPerturb (\S \ref{sec:approach-selfsuper}) with alternative program corruption procedures: Gupta+17, the original DeepFix work that corrupts delimiters or drops variable declarations only (so a subset of our \textbf{Syntax} and \textbf{ID-typo} module),
and Random, a baseline that randomly drops tokens.
We apply DrPerturb, Gupta+17, Random on the DeepFix data to create corresponding training sets (containing 156, 113, 170 types of compiler errors respectively).
We then evaluate repair models trained on each of those training sets on the DeepFix raw test set (Table \ref{tbl:perturb-analysis}).
We find that models trained by DrPerturb  significantly outperform Gupta+17 (+10\% repair rate), suggesting that the diverse set of errors covered by DrPerturb is useful.
Additionally, while Random produces more distinct types of compiler errors than DrPerturb in terms of the number (170 vs 156), models trained by DrPerturb outperform Random by more than 10\%, suggesting that DrPerturb generates a more useful distribution of errors than the Random baseline.

\textbf{Different graph representations.~~}
Table \ref{tbl:graph-analysis} shows an ablation study for the architecture of program-feedback graph.
We find that the edges connecting symbols in source code (``edges among code only'' row), and the edges spanning across source code and a compiler message (``edges across code-feedback only'' row) are equally important,
and the final program-feedback graph (``final'' row) is the most effective. 
We also experimented with a version of our model that uses self-attention (i.e. we consider the complete graph), which we find comparable or slightly less effective (``self-attention'' row). 
This suggests that the most important edges in the graph are those in our program-feedback graph, which connect symbols with semantic correspondence.

\textbf{When is graph \& self-supervision useful?~~}
We study what kinds of compiler errors a program-feedback graph or self-supervised pre-training is most useful for fixing.
Table \ref{tbl:acc_vs_err} shows the breakdown of major compiler errors seen in the SPoC dev set (left), and 
the repair accuracy of our model variants
for each error type (right).  
We used the SPoC data as it exhibited more diverse errors than DeepFix.

We observe that the use of program-feedback graph is particularly helpful for compiler errors such as ``\texttt{@@@ was \!not \!declared}'' and ``\texttt{request \!for \!member \!@@@...}''  (those with bold scores in the ``+\! graph'' column), which typically require analyses of multiple lines of code (recall our example in Fig.\,\ref{fig:task}).
This suggests that a program-feedback graph indeed allows better information flow across source code lines and compiler messages, compared to the baseline sequence model (``base'').
Additionally, 
we observe that self-supervised pre-training improves repair accuracy across most of the error types, but is noticeably helpful for errors that were relatively rare in the \textit{original} training data (e.g. the bottom two), for which the use of a program-feedback graph only helped a little.
This suggests that the extra training examples created in our self-supervision method help mitigate such data scarcity issues in original training data.

\section{Related work and discussion}

\textbf{Graph neural networks.~~}
Graph neural nets (GNN), such as 
graph attention net \citep{velikovi2017graph}, graph convolutional net \citep{kipf2016semisupervised}, graph isomorphism net \citep{how_powerful_gnn} have been shown to be effective for modeling graph-based data.
Several works use GNNs to model the structure of text \cite{Yasunaga&al.17,zhang2018graph} and more recently, source code \cite{learning_to_Allamanis18,generative_code_Brockschmidt19}.
\citet{learning_to_Allamanis18} present {program graph} that augments AST with data flow edges across variables, which
is passed to GNNs to solve the task of variable name prediction.
\citet{generative_code_Brockschmidt19}
build on it and design a graph-based generative model for source code.
\citet{Neural_Attribution} propose a tree convolution model to encode ASTs.
Distinct from the above works, we focus on the problem of program repair, and design the program-feedback graph to represent the dependencies between source code and diagnostic feedback.
Our results show that GNNs can fruitfully represent these program-feedback dependencies for program repair.

\textbf{Self-supervised pre-training.~~}
The idea of using unlabeled data to pre-train neural networks has been shown effective across many fields, including computer vision \citep{vincent_denoising,Erhan_pretrain}, NLP \citep{Peters_2018,devlin2018bert}, graphs \citep{hu2019pretraining}, and programming languages \citep{CodeBERT}. 
Typically, the self-supervised pre-training objective is different from the target task:
For instance, in image recognition, \citet{vincent_denoising} pre-train networks via a denoising autoencoder; in NLP,
\citet{devlin2018bert} pre-train networks via masked language modeling and then fine-tune on a target task such as question answering.
In contrast, our pre-training task \textit{is} the program repair task (our target task), as we prepare the pre-training data by corrupting unlabeled programs and obtaining diagnostic feedback to synthesize program repair examples. 
Additionally, our pre-training task is conditioned on diagnostic feedback, which is a new type of structure from a pre-training perspective and provides better generalization at the test time as we show in \S \ref{sec:results}.

\textbf{Learning program repair.~~}
There is increasing interest in automatic correction of introductory programming assignments \cite{Pu16,Automatic_grading,TRACER}.
DeepFix \citep{DeepFix} is an early work that uses a 
seq2seq model to translate a broken code into fixed one.
RLAssist \citep{Rahul_aaai19}, SampleFix \cite{SampleFix} improve on it by introducing reinforcement learning or better sampling methods.
While these works purely use sequence models,
we propose to use a graph representation of source code and diagnostic feedback
to capture long-range dependencies of symbols across them.

Another line of work learns from labeled datasets of how programmers edit code (e.g. error resolution records) \citep{Defects4J,Chen_2019}.
\citet{DeepDelta} model a Java build error resolution record using seq2seq. 
\citet{Graph2Diff} generalize it to more diverse error types, and propose a repair model that uses the graph structure of AST. 
\citet{Getafix} present a hierarchical clustering algorithm to learn program repair patterns.
While these works rely purely on labeled datasets of program repair,
we propose a self-supervised learning paradigm that leverages a large amount of unlabeled data to create extra training examples for program repair.

Finally, several works focus on repairing specific types of bugs, e.g., variable misuse \citep{Vasic19}, name-based bugs \citep{Pradel_2018} and Javascript bugs \cite{HOPPITY}.
Other works 
focus on modeling program execution \citep{dynamic_emb} or edits \citep{zhao2019neural}.
We refer readers to \citet{repair_review} for a more comprehensive review of automated program repair.

\section{Conclusion}
This paper makes two contributions to program repair from diagnostic feedback.
First, we proposed the program-feedback graph to model the reasoning process in program repair. We find this particularly useful when the repair requires analyzing multiple lines of code.
Second, we introduced a
self-supervised learning paradigm that creates extra program repair examples by corrupting unlabeled programs and obtaining feedback from an evaluator (compiler).
We find this effective for overcoming the scarcity of labeled data for program repair.

While we primarily focus on program repair in this paper, we note that our framework of learning to edit based on feedback is a potentially powerful and more general paradigm with many applications, from learning to edit essays based on written feedback, to learning from users in interactive dialogue, etc. \citep{Liu_2018}. The key is that rather than using a single number reward (e.g. compile or not) as in reinforcement learning, obtaining high bandwidth feedback via diagnostic feedback can be much more informative if we incorporate it effectively, for instance through the use of graph neural networks as we presented in this work.

\section*{Reproducibility}
All code and data are available at \url{https://github.com/michiyasunaga/DrRepair}.
Experiments are available at \url{https://worksheets.codalab.org/worksheets/0x01838644724a433c932bef4cb5c42fbd}.

\section*{Acknowledgments}
We thank John Hewitt, Mina Lee, Sumith Kulal, Pang Wei Koh, Robin Jia, Ananya Kumar, Ruiqi Zhong, and anonymous reviewers for insightful feedback and discussions.
This work was supported in part by NSF CAREER Award IIS-1552635, a PECASE Award, and an Amazon Research Award.

\bibliography{main}
\bibliographystyle{icml2020}





\end{document}

%% file: tbl_eval_deepfix_v2.tex
\begin{table*}[t]

\definecolor{myydarkblue}{HTML}{1155cc} 
\definecolor{myorange}{HTML}{ff8100}
\definecolor{mydarkorange}{HTML}{bf6100}
\definecolor{myviolet}{HTML}{a805c4}
\definecolor{myred}{HTML}{af1503}

\vskip -0.04in
\centering
\renewcommand{\arraystretch}{1.04}
\scalebox{0.85}{\textbf{Evaluation on DeepFix data}}\\[0.5mm]
\scalebox{0.85}[0.85]{
\begin{small}
\begin{tabular}{l||cc|c}
\Xhline{2\arrayrulewidth}
\hspace{-0mm}\multirow{2}{*}{\textbf{Repair Model}} &
\scalebox{0.95}{\textbf{Single Localize}}
& \scalebox{0.95}{\textbf{Single Repair}} 
& ~~\textbf{Full Repair}~~\vrule width 0pt height 11.5pt depth 5pt
\\[-0.6mm]
& 
~\scalebox{0.85}[0.85]{(Our \textcolor{myydarkblue}{\textbf{synthetic}} dev)}\vrule width 0pt height 0pt depth 5pt ~
& ~\scalebox{0.85}[0.85]{(Our \textcolor{myydarkblue}{\textbf{synthetic}} dev)}~
& ~~\scalebox{0.9}[0.9]{(\textcolor{myviolet}{\textbf{DeepFix raw test}})}~~\\
\hline
\hspace{-0mm}\scalebox{0.96}[1]{DeepFix \scalebox{0.9}{~\hspace{3.1mm}\citep{DeepFix}}} \vrule width 0pt height 10pt depth 0pt & 
- & - & \hspace{1.5mm}27.0$^{*}$ \\
\hspace{-0mm}\scalebox{0.96}[1]{RLAssist \scalebox{0.9}{~\hspace{2.1mm}\citep{Rahul_aaai19}}}  & 
- & -  & \hspace{1.5mm}26.6$^{*}$\\
\hspace{-0mm}\scalebox{0.96}[1]{SampleFix \scalebox{0.9}{~\citep{SampleFix}}}  & 
- & - & \hspace{1.5mm}45.3$^{*}$\\[0.3mm]
\hdashline[2pt/1.5pt]&&&\\[-2.7mm]
\hspace{-0mm}\scalebox{0.96}[1]{\textbf{Our} base (no compiler message)}  & 
95.0 & 70.8 & \hspace{1.5mm}34.0$^{*}$\\
\hspace{-0mm}\scalebox{0.96}[1]{\textbf{Our} base} & 97.1 & 70.9 & 62.5\\
\hspace{-0mm}\scalebox{0.96}[1]{\textbf{Our} base + graph} & 97.9 & 74.8 & 66.4\\
\hspace{-0mm}\scalebox{0.96}[1]{\textbf{Our} base + graph + pre-train\, \scalebox{0.9}{(DrRepair)}}\vrule width 0pt height 0pt depth 3.5pt & 
\textbf{98.9} & \textbf{80.2} & \textbf{68.2} \\
\bottomrule \multicolumn{4}{l}{}\\[-3mm]
\end{tabular}
\end{small}}
\vskip -0.09in
\caption{Performance of our repair model and prior work on \textbf{DeepFix} data. We report the single step error localization\,/\,repair accuracy (\%) on our \textbf{synthetic dev} set (column 2-3), and the {full repair success rate} (\%) on DeepFix \textbf{raw test} set (column 4).
``{DrRepair}'' refers to our full model, which
outperforms prior work by significant margins. ~
\scalebox{0.8}{(*) compiler messages not used.}
}\vspace{-5mm}
\label{tbl:deepfix-eval}
\end{table*}

%% file: tbl_eval_spoc_dev.tex

\begin{table}[t]
\vskip 0.05in
\centering
\scalebox{0.85}{\textbf{Ablation on SPoC dev}}\\[1.3mm]
\scalebox{0.85}{
\begin{small}
\begin{tabular}{l|cc}
\Xhline{2\arrayrulewidth}
\hspace{-1mm}\multirow{3}{*}{\textbf{Repair Model}}\vrule width 0pt height 11pt depth 5pt & \scalebox{0.99}{\textbf{Single}}\vrule width 0pt height 10.5pt depth 0pt
& \scalebox{0.99}{\textbf{Single}}
\\[-0.7mm]
& \scalebox{0.99}{\textbf{Localize}}  & \scalebox{0.99}{\textbf{Repair}}\\[0.5mm]
& \!\scalebox{0.85}[0.85]{(SPoC dev)}\!\vrule width 0pt height 0pt depth 4.5pt & \!\scalebox{0.85}[0.85]{(SPoC dev)}\!\! \\
\hline
\hspace{0mm}\scalebox{0.96}[1]{Our base} \vrule width 0pt height 10pt depth 0pt & 92.0  & 48.6 \\
\hspace{0mm}\scalebox{0.96}[1]{Our base + graph} & 
93.1  & 53.0\\
\hspace{0mm}\scalebox{0.96}[1]{Our base + graph + pre-train\, \scalebox{0.9}{(DrRepair)}}\! & 
\textbf{94.9} & \textbf{56.2}\\[0.7mm]
\hdashline[2pt/1.5pt]\\[-2.5mm]
\hspace{-1.5mm}\textbf{\scalebox{0.93}{If use pseudocode}}   &  &   \\
\hspace{0mm}\scalebox{0.96}[1]{Our base}    & 
93.2  &  {65.2} \\
\hspace{0mm}\scalebox{0.96}[1]{Our base + graph + pre-train\, \scalebox{0.9}{(DrRepair)}}\! & 
\textbf{96.1} & \textbf{68.0}  \\
\bottomrule
\end{tabular}
\end{small}}
\vskip -0.07in
\caption{Performance of our repair model on \textbf{SPoC} data. We measure the {single step} error localization\,/\,repair accuracy (\%) on the {SPoC Dev} set. ``{DrRepair}'' refers to our full model.
} \vspace{-4mm}
\label{tbl:spoc-dev}
\end{table}

%% file: tbl_eval_spoc_test.tex
\begin{table}[t]
\vskip 0.05in
\centering
\scalebox{0.85}{\textbf{Evaluation on SPoC test}}\\[1.3mm]
\scalebox{0.85}{
\begin{small}
\begin{tabular}{l||cc|cc}
\Xhline{2\arrayrulewidth}
\hspace{-1mm}\multirow{2}{*}{\textbf{Synthesis Method}} &
\multicolumn{2}{c|}{\textbf{SPoC TestP}\vrule width 0pt height 12.5pt depth 5pt} & \multicolumn{2}{c}{\textbf{SPoC TestW}} \\
\vrule width 0pt height 0pt depth 4.5pt & \!\scalebox{0.93}{$B$=10}\! & \!\!\scalebox{0.93}{$B$=100}\! & \!\scalebox{0.93}{$B$=10}\! & \!\!\scalebox{0.93}{$B$=100}\!\!\!\\
\hline
\hspace{-1mm}{\scalebox{1}{Baselines}}\vrule width 0pt height 10pt depth 4pt &&&& \\
\hspace{3mm}Top\,1 (no search) & 17.8 & 17.8 & 30.7 & 30.7 \\
\hspace{3mm}Best first search & 26.5 & 32.5 & 42.5  & 51.0\\[0.6mm]
\hspace{-1mm}{\scalebox{1}{Prior best}}~ \scalebox{0.9}{\cite{spoc2019}} & 
{28.4} & {34.2}  & 44.4  &  53.7\\[0.6mm]
\hdashline[2pt/1.5pt] &&&&\\[-2.1mm]
\hspace{-1mm}\textbf{Our} \scalebox{0.93}{DrRepair}   & 30.2 & 37.5 & 46.6 & 55.9\\
\hspace{-1mm}\textbf{Our} \scalebox{0.93}{DrRepair  \scalebox{0.9}[0.95]{w/} pseudocode}\!\!  & \textbf{31.4} & \textbf{38.5} & \textbf{48.0} & \textbf{57.0}\\[0.3mm]
\bottomrule
\end{tabular}
\end{small}}
\vskip -0.07in
\caption{Program \textbf{synthesis success rate} (\%) at search budgets $B$ on the \textbf{SPoC Test} sets.
Our search method equipped with DrRepair consistently outperforms the previous best in all settings.}\vspace{-4mm}
\label{tbl:spoc-test}
\end{table}

%% file: tbl_perturb_analysis.tex
\begin{table}[t]
\vskip -0.0in

\centering
\scalebox{0.85}{
\begin{small}
\begin{tabular}{l|cccc}
\Xhline{2\arrayrulewidth}
\hspace{-1mm}\multirow{3}{*}{\textbf{Repair Model}} &
\multicolumn{3}{c}{\textbf{Corruption Procedure}\vrule width 0pt height 11.5pt depth 6pt} \\
\vrule width 0pt height 0pt depth 5.5pt & \scalebox{0.93}[0.96]{DrPerturb}\! & \multirow{1}{*}{\!\scalebox{0.93}[0.96]{Gupta+17}\!} & \multirow{1}{*}{\!\scalebox{0.93}[0.96]{Random}\!\!\!} \\[-1.7mm]
& \scalebox{0.85}{(\textbf{Ours})} & & \\[0.3mm]
\hline
\hspace{-1mm}\scalebox{0.95}[1]{{Our} base (no compiler feedback)}\vrule width 0pt height 10.5pt depth 0pt\!  & 
\textbf{34.0} & 24.2 & 30.1\\
\hspace{-1mm}\scalebox{0.95}[1]{{Our} base} & 
\textbf{62.5} & 50.5 & 49.4
\\
\hspace{-1mm}\scalebox{0.95}[1]{{Our} base + graph} & 
\textbf{66.4} & 54.5 & 53.0
\\
\bottomrule
\end{tabular}
\end{small}}
\vskip -0.07in
\caption{
\textbf{Effect of different program corruption procedures}. We train repair models using different program corruption methods (our DrPerturb, \citet{DeepFix}'s, and random token dropout), and evaluate the trained models on the DeepFix raw test set (full repair rate \%).
\textbf{Bold score} indicates the best corruption algorithm. 
}\vspace{-3mm}
\label{tbl:perturb-analysis}
\end{table}

%% file: tbl_graph_analysis.tex
\begin{table}[t]
\vskip 0.0in
\centering
\scalebox{0.85}{
\begin{small}
\begin{tabular}{l|c}
\Xhline{2\arrayrulewidth}
\hspace{-1mm}\textbf{Repair Model} \vrule width 0pt height 11pt depth 5pt & \!\textbf{Full Repair}\!\!\\
\hline
\hspace{-1mm}\scalebox{0.95}[1]{Our base} \vrule width 0pt height 10pt depth 0pt & 62.5  \\
\hspace{-1mm}\scalebox{0.95}[1]{Our base + graph (edges among code only)} & 64.8  \\
\hspace{-1mm}\scalebox{0.95}[1]{Our base + graph (edges across code-feedback only)}\! & 64.9\\
\hspace{-1mm}\scalebox{0.95}[1]{Our base + graph (final)} & \textbf{66.4}\\
\hspace{-1mm}\scalebox{0.95}[1]{Our base + self-attention} &  66.0\\
\bottomrule
\end{tabular}
\end{small}}
\vskip -0.07in
\caption{\textbf{Comparison of different graph architectures}. We evaluate the models on the Deepfix raw test set (full repair rate \%).
} \vspace{-3mm}
\label{tbl:graph-analysis}
\end{table}